# The Micro-Randomized Trial for Developing Digital Interventions: Experimental Design Considerations

Ashley E. Walton[1]

Dartmouth College

Linda M. Collins[2]

Pennsylvania State University

Predrag Klasnja[3]

University of Michigan

Inbal Nahum-Shani[4]

University of Michigan

Mashfiqui Rabbi[5]

Harvard University

Maureen A. Walton[6]

University of Michigan

Susan A. Murphy[7]

Harvard University

# Author Note

[1]Ashley E. Walton, Department of Philosophy, Dartmouth College. Dr. Walton was supported by National Institutes of Health grants RF1MH117813, P50DA039838, R01AA023187, and U54EB020404

[2]Linda M. Collins, The Methodology Center and Department of Human Development & Family Studies, The Pennsylvania State University. Dr. Collins was supported by National Institutes of Health grants P50DA039838, R01AA022931, P01CA180945, and R01DA040480.

[3]Predrag Klasnja, School of Information, University of Michigan. Dr. Klasnja was supported by National Institutes of Health grants R01HL125440, U01CA229445, and R01LM013107.

[4]Inbal Nahum-Shani, Institute for Social Research, University of Michigan. Dr. Nahum-Shani was supported by National Institutes of Health grants: U01 CA229437, R01 DA039901, R01 CA224537, R01 AA026574, R01 DK108678; and by Patient-Centered Outcomes Research Institute grant PCS-2017C2-7613

[5]Mashfiqui Rabbi, Department of Statistics, Harvard University. Dr. Rabbi was supported by National Institutes of Health grants P50DA039838, R01AA023187, U01CA229437, and U54EB020404.

[6]Maureen A. Walton, Department of Psychiatry and Injury Prevention Center, University of Michigan. Dr. Walton was supported by funding from the University of Michigan Injury Prevention Center (CDC R49CE002099).

[7]Susan A. Murphy, Departments of Statistics and Computer Science, Harvard University. Dr. Murphy was supported by National Institutes of Health grants P50DA039838, R01AA023187, U01CA229437, UG3DE028723, and U54EB020404.

The corresponding author is Susan A. Murphy, Science Center 400 Suite, Harvard University, One Oxford Street, Cambridge, MA 02138-2901

The authors thank Amanda Applegate for helpful comments.

**Abstract**

Just-in-time adaptive interventions (JITAIs) are time-varying adaptive interventions that can deploy a high intensity of adaptation; in other words, use frequent opportunities for the intervention to be adapted—weekly, daily, or even many times a day. This high intensity of adaptation is facilitated by the ability of digital technology to continuously collect information about an individual's current context and make treatment decisions adapted to this information. The micro-randomized trial (MRT) has emerged for use in informing the construction of JITAIs. MRTs operate in, and take advantage of, the rapidly time-varying digital intervention environment. MRTs can be used to address research questions about whether and under what circumstances particular components of a JITAI are effective, with the ultimate objective of developing effective and efficient components. The purpose of this article is to clarify why, when, and how to use MRTs; to highlight elements that must be considered when designing and implementing an MRT; and to discuss the possibilities this emerging optimization trial design offers for future research in the behavioral sciences, education, and other fields. We briefly review key elements of JITAIs, and then describe three case studies of MRTs, each of which highlights research questions that can be addressed using the MRT and experimental design considerations that might arise. We also discuss a variety of considerations that go into planning and designing an MRT, using the case studies as examples.

# The Micro-Randomized Trial for Developing Digital Interventions:
# Experimental Design Considerations

Just-in-time adaptive interventions (JITAIs), which are receiving a tremendous amount of attention in many areas of behavioral science (Nahum-Shani et al., 2018), are time-varying adaptive interventions delivered via digital technology. JITAIs use a high intensity of adaptation; in other words, there are frequent opportunities for the intervention to be adapted—weekly, daily, or even many times a day. This high intensity of adaptation is facilitated by the ability of digital technology to continuously collect information about an individual's current context and make treatment decisions adapted to this information. A JITAI may constitute an entire digital intervention, or it may be one of multiple components in an intervention.

JITAIs are typically provided as "push" intervention components, in which the intervention content is delivered to individuals via system-initiated interactions, such as push notifications via a smartphone or smart speaker and haptic feedback on a smart watch. In addition to push components, digital interventions may also include "pull" intervention components, which provide content that individuals can access any time, at will. The effectiveness of pull components rests on the assumption that the individual will recognize a need for support and actively decide to access the pull component (Nahum-Shani et al., 2018). By contrast, push intervention components do not require that the participant recognize when support is needed—or even remember that support is available on the digital device. Instead, sensors on smart devices continuously monitor an individual's context, enabling intervention content to be delivered when needed, irrespective of whether or not the individual is aware of this need.

Push components are a potentially powerful and versatile intervention tool, but they have an inescapable down side: they may interrupt individuals as they go about their daily lives. If these interruptions become overly burdensome or irritating, there is a risk of disengagement with the intervention (Rabbi et al., 2018). Furthermore, repeated notifications used to provide push interventions can lead to habituation: the reduced level of responsiveness resulting from frequent stimulus exposure. When habituation occurs, the individual's attention to the push stimulus deteriorates, possibly to the point at which the individual no longer notices the stimulus. Thus, it is good practice to limit content delivered by push intervention components to the minimum needed to achieve the desired effect. This can be accomplished by strategically developing JITAI push components that deliver content only in the contexts in which they are most likely to be effective and eliminating any low-performing push components that do not result in enough behavior change to compensate for the attention required by the participant.

The above considerations justify optimizing JITAI push intervention components prior to evaluation in an RCT and subsequent implementation. The micro-randomized trial (MRT; Klasnja et al., 2015; Liao, Klasnja, Tewari, & Murphy, 2016) has emerged for use in informing the construction of JITAIs. MRTs operate in, and take advantage of, the rapidly time-varying digital intervention environment. MRTs can be used to address research questions about whether push intervention components are effective and in which time-varying states they are effective, with the ultimate objective of developing effective and efficient JITAI components.

The purpose of this article is to clarify why, when, and how to use MRTs; to highlight elements that must be considered when designing and implementing an MRT; and to discuss the possibilities this emerging optimization trial design offers for future research in the behavioral sciences, education, and other fields. This article lies in between the high-level overview of the MRT for health scientists provided by Klasnja et al (2015) and the statistical paper, primarily focused on methods for sample size calculations, by Liao et al. (2016). This article provides a more in depth and updated discussion of considerations that inform the design of an MRT, based on our experiences conducting MRT studies, such as the case studies described below. Analysis of data produced by an MRT is discussed in a companion article (Qian et al., under review).

**Elements of Just-in-Time Adaptive Intervention Components**

Although this article is primarily about experimental design as it pertains to MRTs, to consider MRTs it is necessary to consider the design elements of JITAI intervention components. (As described below, these design elements may themselves be considered components of a JITAI if they are separated out for study in an optimization trial; see Collins, 2018. Thus, in this article we are using the term component broadly.) While reading this article one should be mindful of the distinction between experimental design and intervention design. For example, the MRT is a type of experimental design, and the JITAI is a type of intervention design. Here we briefly review key elements of JITAI design (in italics) that will be discussed in more detail later.

Like most interventions, digital interventions are typically developed with the objective of improving one or more long-term health outcomes, which we will call *distal outcomes*. The strategy for improving the distal outcomes involves the provision of one or more *intervention components*. We focus on JITAIs here, but we emphasize that digital interventions may be a mix of JITAI components and other types of components, such as fixed components. Each JITAI may have two or more *intervention component options* (e.g., deliver an SMS message saying "The weather forecast says it will be a beautiful day for a walk!" or do not deliver an SMS message). Ideally the components and component options of all evidence-based interventions are conceived based on a conceptual model that has been informed by theory and empirical evidence (Collins, 2018; Nahum-Shani et al., 2018). A conceptual model specifies how each component of an intervention is designed to affect distal outcomes via one or more specific mediators, or *proximal outcomes*, that are part of the hypothesized causal process through which the intervention is intended to work. These proximal outcomes may, in turn, directly affect the distal outcomes; or they may be part of a longer causal chain in which proximal outcomes affect subsequent proximal outcomes until the distal outcome is reached.

The high intensity of adaptation that characterizes JITAI components means interventions may be varied frequently by providing individuals with different component options at prespecified times called *decision points*. *Observations of context*—data on the individual's context, such as aspects of the individual's current external and intrapersonal environments and the individual's history, are often used to tailor the content of the different options of JITAI component. At each decision point, a *decision rule* specifies which option to provide based wholly or partially on observations of context available to the smart device.

**Introduction to the MRT**

The MRT is an optimization trial that can be used to assess the performance of JITAI intervention components and component options. For example, an MRT can be used to address questions about selection of elements of the design of the JITAI, such as in which time-varying context each component option is best and in which time-varying context it is best to provide no intervention. In short, *an MRT is used to optimize the JITAI decision rules, with the ultimate goal of developing an effective and efficient JITAI*. Below we review the essential features of an MRT.

A *factor* is a variable that is experimentally manipulated in the MRT. An MRT can be used to investigate one or more factors, each corresponding to an intervention component, and each having *levels* corresponding to that component's options (see Collins, 2018). As will be shown in the case studies, not all JITAI components are necessarily randomized in a single MRT. Components that are not randomized in an optimization trial, that is, are delivered as usual to all participants, are called constant[1] components (Collins, 2018) to distinguish them from those that are experimentally manipulated.

*Decision points* are pre-determined times at which it might be useful to deliver a component. In an MRT the decision points may be specific to a particular JITAI component and therefore factor-specific; that is, each factor has its own set of decision points. The discussion of the case studies highlights that this specification must be made carefully because the frequency and timing of decision points can be critical for intervention effectiveness.

Randomization facilitates the estimation of causal effects. A primary rationale for randomization in any experimental design is that it enhances balance in the distribution of unobserved variables across groups receiving different treatments, reducing the number of alternative explanations for why a group assigned one treatment has better outcomes than a group assigned a different treatment. In an MRT participants are *sequentially randomized* to the different levels of each factor at hundreds or even thousands of decision points over the course of the experiment. These repeated randomizations in an MRT play essentially the same role: the randomization enhances balance in the distribution of unobserved factors between decision points assigned to different intervention options. This enables the investigator to use the results as a basis for answering causal questions concerning whether a component option has the desired effect on the proximal outcome and whether this effect varies with time and context. As will be described in the case studies, an optimization trial can include both micro-randomized and baseline-randomized factors.

---

[1] The term "constant" means the options of a component are not being manipulated in the trial; constant does not refer to time-invariant. A constant component may or may not be time-varying and may or may not be adaptive.

*Randomization probabilities* are the pre-specified probabilities of randomly assigning participants to the levels of a factor (i.e., the options of a particular component). As will be shown in the case studies, the randomization probabilities associated with the levels of a factor in an MRT (unlike most classical factorial experiments) are not necessarily equal. For example, because participant burden is an important consideration when selecting randomization probabilities, burden may be reduced strategically by assigning larger randomization probabilities to less burdensome levels.

*Observations of context* are variables of practical or scientific interest recorded at a particular decision point, or summaries of variables observed prior to the decision point. Observations of context may be gathered by means of self-report measures; recorded as part of the treatment (e.g., amount of support received in the past week); or captured by mobile devices (e.g., location, weather, movement), wearable sensors (e.g., heart rate, step count), and other electronic devices (e.g., wireless scales participants use to weigh themselves). In the design of an MRT, observations of context play two distinct roles. First, observations of context may serve as *availability conditions* that determine whether or not it is appropriate to deploy a particular factor level (component option) at a particular decision point. The scientific team may decide a priori that in certain contexts deployment of a component option would be inappropriate on scientific grounds or because it would be potentially irritating to or unsafe for the participant. In this case the participant is considered *unavailable* and the "do nothing" option is automatically selected (i.e., no treatment randomization occurs). Second, observations of context may be collected in an MRT because they are potential moderators that can be used to identify which option performs best in which context, thus informing the development of decision rules in the optimized JITAI. See the companion paper (Qian et al., submitted) for examples of moderation analyses.

## Case Studies

In this section, we review three case studies of optimization trials involving MRTs; each case study highlights research questions that can be addressed using the MRT and experimental design considerations that might arise. Case Study 1 describes HeartSteps. The goal of the HeartSteps intervention is to increase physical activity among sedentary individuals (Klasnja et al., 2015; Klasnja et al., 2019). This case study will be used to illustrate the essential features of an MRT reviewed above. Case Study 2 describes the Substance Abuse Research Assistant (SARA), an app to collect data on substance use by at-risk adolescents and young adults. This case study highlights how MRTs can be used to optimize a JITAI aimed at improving data collection by an app (Rabbi et al., 2018). Case Study 3 describes BariFit, an intervention to support weight maintenance for individuals who have undergone bariatric surgery (Ridpath, 2017). This case study demonstrates how it can be appropriate to include both baseline-randomized factors and micro-randomized factors in an optimization trial. Figures giving gestalt overviews of each study can be found at https://www.methodology.psu.edu/ra/adap-inter/mrt-projects/#proj.

**Case Study 1: HeartSteps**

The long-range objective of the HeartSteps project is to improve the outcome of heart health in adults by helping individuals with heart disease achieve and maintain recommended levels of physical activity. Physical activity is known to decrease cardiovascular risks, yet only one in five adults in the U.S. meets the guidelines for the number of minutes of physical activity recommended per week (Centers for Disease Control & Prevention, 2014). The HeartSteps MRT was designed to optimize two push components for improving physical activity. Optimization of these two components was intended to improve the distal outcome of average daily step count over the 42-day study.

**Intervention components and component options**. The initial version of the HeartSteps intervention included a number of components. Here we focus on the two push components investigated by the MRT.

The first component, Activity Suggestions, consisted of contextually tailored suggestions intended to increase opportunistic physical activity. Finding time to exercise can be challenging for adults with full-time work schedules. However, individuals can still experience health benefits by engaging in opportunistic physical activity, in which brief periods of movement or exercise are incorporated into the daily routine (Centers for Disease Control & Prevention, 2018). Activity suggestions were provided as push notifications delivered to the participant's smartphone. There were three different options for this intervention component: participants could receive either a suggestion with a walking activity that took 2-5 minutes to complete, an anti-sedentary suggestion (instructing brief movements) that took 1-2 minutes to complete, or no suggestion. HeartSteps illustrates that intervention components can, and in fact often do, include an option of "do nothing."

The second component, Planning Support, consisted of support for planning how to be active the next day. This component was motivated by scientific evidence demonstrating that engaging in a behavior can require less effort when individuals develop plans that articulate exactly when, where, and how they will be more likely to engage in that behavior (Gollwitzer, 1999). This component had three options. Participants could receive either a prompt asking them to select a plan from a list containing their own past activity plans (structured planning); a prompt asking them to type their plan into a text box (unstructured planning); or no prompt.

The HeartSteps intervention also included several constant components, for example a self-monitoring component that assisted participants in tracking their activity and a library of previously sent activity suggestions. Thus HeartSteps illustrates how it is possible to select only a subset of the components in a digital intervention for experimentation in an MRT, while others are delivered as usual to all participants.

**Decision points**. Originally the investigative team planned to have a decision point every minute of the waking day in order to allow the Activity Suggestions component to arrive in real time. However, prior data on employed individuals indicated that the greatest within-person variation in step counts occurred around the morning commute, lunch time, mid-afternoon, evening commute, and after dinner times (Klasnja et al., 2015), indicating that at these times there is greater potential to increase activity. In HeartSteps, the actual times of these decision points were specified by each individual at the start of the study, and thus varied by participant. Viewing these five times as times during the day at which individuals are most likely to be responsive because they would have opportunities to walk, the team decided to use these five times as the decision points for the Activity Suggestions component. On the other hand, because the Planning Support component involved planning the following day's activity, the natural choice of a decision point was every evening at a time specified by each participant at the beginning of the study.

**Observations of Context and Availability Conditions**. HeartSteps illustrates how observations of context can inform the content of an intervention component: the suggestion in the Activity

Suggestions component was tailored according to the participant's current location, current weather conditions, time of day, and day of the week. This was intended to make the suggestions immediately actionable and more easily incorporated into a participant's daily routine (Rabbi et al., 2018). In HeartSteps, availability conditions pertained primarily to the Activity Suggestions component. In HeartSteps, participants could be considered unavailable to be sent an activity suggestion for several reasons. First, they were considered unavailable if sensors on the phone indicated that they might be operating a vehicle. Second, because all of the contextually tailored activity suggestions asked participants to walk, the research team felt it would be inappropriate to send one of these suggestions if sensors indicated that the participant is already walking or running or just finished an activity bout in the previous 90 seconds. Third, participants could turn off the activity notifications for 1, 2, 4, or 8 hours, to enable them to exert some control over the delivery of the suggestions.

In addition to the observations of context described above, a number of additional observations of context were collected not for use in adapting delivery of intervention content during the study, but rather for exploratory moderation analyses after study completion. For example, current location and weather, and number of days in the study were potential moderators for use in understanding when and in which context it is best to provide an activity suggestion in a future version of HeartSteps.

**The HeartSteps Optimization Trial.**

*Factors in the Experiment.* This 42-day MRT included two factors, one corresponding to the Activity Suggestions component and one corresponding to the Planning Support component. Each factor had three levels, corresponding to the component options.

*Measures of Proximal Outcomes.* The initial version of HeartSteps focused primarily on increasing daily physical activity through walking; therefore, step count was used to form the proximal outcomes. Minute-level step counts were passively recorded using a wristband activity tracker. The proximal outcome for the Planning Support component was the total number of steps taken on the subsequent day because the planning was for the next day's physical activity. Deciding how to operationalize the proximal outcome for the Activity Suggestions component was more challenging. A 5- or even 15-minute duration for the total step count following a decision point would be too short, as the individual might not have enough time to act on the suggestion. On the other hand, since some activity suggestions only asked participants to engage in a short bout of activity to disrupt their sedentary behavior, the research team was concerned that a proximal outcome that was longer, like an hour, would be too noisy to detect the impact of the anti-sedentary suggestions. Ultimately, the team settled on the total number of steps taken in the 30 minutes following each decision point.

*Primary and Secondary Research Questions.* The HeartSteps MRT was conducted to address the following primary research question:

1. Is there an overall effect of Activity Suggestions? On average across time, does providing the contextually tailored activity suggestions increase physical activity in the 30 minutes after the suggestion is delivered, compared to no suggestion?
    a. If so, does the effect deteriorate with time (day in study)?

Examples of secondary research questions include

2. Is there an overall effect of Planning Support? On average across time, does pushing a daily activity planning support prompt increase physical activity the following day compared to no prompt?

        a. If so, does the effect deteriorate with time (day in study)?

3. Concerning the Activity Suggestions factor: On average across time, is there an overall difference between the walking activity suggestion and the anti-sedentary activity suggestion on the subsequent 30-minute step count?

4. Concerning the Planning Support factor: On average across time, is there a difference between the structured, lower-burden option and the unstructured, higher-burden option on the next day's physical activity?

Additional exploratory analyses were planned with the objective of understanding whether context moderated the effects of either of the factors. For example, the team was interested in whether location moderated the effectiveness of the Activity Suggestions component and whether day of week moderated the effectiveness of the Planning Support component. These moderation analyses are for use in developing decision rules informing the delivery of the components (e.g., perhaps the activity suggestion is effective only when the individual is at home or work, indicating that the next iteration of HeartSteps should deliver the activity suggestions only in these locations).

***Randomization***. Figure 1 provides a schematic to illustrate the randomization for the Activity Suggestions factor. During pilot testing to prepare for the HeartSteps MRT, the randomization probabilities for the Activity Suggestions component were initially selected so as to deliver an average of two activity suggestions per day across the five decision points. Two suggestions per day was deemed the appropriate frequency to minimize burden and reduce the risk of habituation. However, it became clear that, on average, approximately one suggestion per day was never seen because individuals left their phones in a bag or coat pocket. The investigators decided that to increase the likelihood of at least two activity suggestions being seen, it was necessary to deliver more than two suggestions. Therefore, the randomization probabilities were adjusted before beginning the MRT so that, on average, three activity suggestions would be delivered per day. As Figure 1 illustrates, the randomization probabilities assigned to the options of the Activity Suggestions component were walking activity suggestion, 0.3; anti-sedentary suggestion, 0.3; no suggestion, 0.4. Thus the probability of receiving a suggestion (as opposed to no suggestion) was 0.6, resulting in an expected average of three suggestions delivered per day, with two out of the three seen per day.

Because for the Planning Support component there was one decision point per day, in the evening, at a convenient time selected by the participant, participants were considered always available. Thus, availability was not a consideration in randomization for this factor. The randomization probabilities assigned to the options of the Planning Support component were structured planning prompt, 0.25; unstructured planning prompt, 0.25; no prompt, 0.5.

**Case Study 2: Substance Abuse Research Assistant (SARA)**

A combination of individual, social, and environmental factors can increase the risk of youth developing a substance use disorder (Rabbi et al., 2018). Yet little is understood about the temporal processes that underlie this development and how and when these factors might be targeted for intervention (Rabbi et al., 2018). The Substance Abuse Research Assistant (SARA) is a mobile application for collection of data about the time-varying correlates of substance use among youth reporting recent binge drinking and/or marijuana use. Every day at 6pm, SARA prompted participants with a survey to report their feelings and experiences for that day, including their perceptions of stress, mood, loneliness, and amount of free time over the day; participants could fill out the survey anytime between 6 pm and midnight. On Sundays, the survey included additional questions about their substance use that week, such as frequency of use, motivation to use, and whether they intended to avoid substance use in the upcoming week. Participants were also asked every day to complete two active tasks—one to measure their spatial memory and one to measure their reaction time. These measures were included due to their potential relationship to the effects of substance use (Celio et al., 2014; Schoeler & Bhattacharyya, 2013).

The prospect of using mobile technology for this kind of data collection is exciting. Most youth own smartphones that can be used to capture information continuously throughout their everyday lives, so mobile technology can be a powerful tool to collect data on the moment-to-moment influences on their substance use. However, this technology is useless if youth will not enter data. Establishing and maintaining engagement with data collection apps is a challenge, and achieving consistent rates of self-report often requires costly incentives or staff interactions. For these reasons, the aim of the SARA MRT was to examine several engagement components designed to sustain or improve rates of self-reporting via the SARA app. The objective of the engagement components was to affect the distal outcome of overall survey and active task completion during the 30-day study.

**Intervention Components and Intervention Component Options**. Here, we focus on four components aimed at increasing and maintaining engagement that were examined in the SARA MRT (Rabbi et al., 2018). The Reciprocity Notification component consisted of a push notification sent 2 hours before the daily data collection period (6pm to 12 midnight). There were two component options: a reciprocity notification containing an inspirational message in the form of youth-appropriate song lyrics or a celebrity quote, or no notification. This component was motivated by behavioral science indicating that delivering an incentive may facilitate participants' desire to return the favor, in this case by completing the survey and active tasks

MRT FOR DEVELOPING DIGITAL INTERVENTIONS: DESIGN                                    19(Blau, 1968; Cialdini et al., 1975; Gouldner, 1960). The Reminder Notification component was a daily push notification delivered at the beginning of the data collection period. The component options were a simple reminder or a reminder that contained a persuasive message. The persuasive messages included content to encourage self-report completion by reminding participants of the rewards they could receive and the short time it required to complete the survey and active tasks. The Post-Survey Reinforcement component was delivered immediately after completion of the survey, to serve as a positive reinforcement for the survey-completion behavior (Ferster & Skinner, 1957/1997; Reynolds, 1975). The two intervention options were a notification containing a reward in the form of a meme or gif or no post-survey reinforcement. Only individuals who completed the survey were eligible to receive the reward. The Post-Active-Task Reinforcement component, delivered immediately after completion of the active tasks, was based on research suggesting that individuals desire knowledge about themselves and their personal abilities and are interested in feedback regarding their behavior (Singh et al., 2016). This component had two options. One option was a notification containing a life insight, which were visualizations of participants' self-reported data from the past seven days (e.g., visualizations of reported daily stress); the other option was no notification. Only individuals who completed the active tasks were eligible to receive the reinforcement.

      The SARA mobile application contained several engagement strategies that were treated as constant components. For example, SARA included a growing virtual aquarium that was progressively filled with fish as participants earned points for completing data collection (see Rabbi et al., 2018). Completion of the survey and active tasks would "unlock" and add a new fish to the aquarium on average once per day. Participants could earn additional treasures for streaks in self-reporting and graduate to different "levels" where the aquarium tank turned into a

sea environment. This game-like structure, designed to support a positive experience, served as the base engagement strategy for SARA. Because this component did not involve push notifications, making it unlikely to interrupt or burden participants, and there were no high-priority scientific questions concerning this component, it was decided not to experiment on it in this optimization trial.

**Decision Points.** Each of the four components had at most one decision point per day. For the Reciprocity Notification component decision points were daily at 4 pm, after school but before the data collection period. Reminder Notification decision points were daily at 6 pm, the start of the data collection period. Post-Survey Reinforcement decision points immediately followed completion of the survey. Similarly, Post-Active-Task Reinforcement decision point immediately followed completion of the active tasks.

**Availability Conditions.** For the Reciprocity Notification and Reminder Notification components there were no availability conditions, as the content of the notifications were not tailored to the participant's current context and because they were programmed to be available for participants to read any time between delivery and midnight. For the Post-Survey Reinforcement and Post-Active-Task Reinforcement components, participants who did not complete the survey or the tasks, respectively, were considered unavailable.

**The SARA Optimization Trial.**

*Factors in the Experiment*. This 30-day MRT included four factors, corresponding to the intervention components: Reciprocity Notification, Reminder Notification, Post-Survey Reinforcement, and Post-Active-Task Reinforcement. Each factor had two levels, corresponding to the component options.

*Measures of Proximal Outcomes*. Because both the Reciprocity Notification and Reminder Notification components were intended to impact that evening's data collection, the proximal outcome for both was whether or not participants completed either the survey or the active tasks on *that same day*. By contrast, both the Post-Survey Reinforcement and Post-Active-Task Reinforcement were intended to increase data collection on the following day; therefore, the proximal outcome for these components was whether or not participants completed the survey and/or the active tasks on *the next day*.

*Observations of Context* were collected primarily for use in analyses of data from the MRT, including day of the week. Other observations included the prior day's self-reporting, as well as use of the SARA app unrelated to survey or active tasks completion. In addition, participants had the option of rating the memes and life insights with a "thumbs up" or "thumbs down."

*Primary and Secondary Research Questions*. The SARA MRT was conducted to address the following primary research questions:

1. Is there an overall effect of Reciprocity Notification? On average across time, does providing an inspirational message two hours before data collection result in increased completion of the daily survey and/or active tasks on that same day compared to no inspirational message?

2. Is there an overall effect of Post-Survey Reinforcement? On average across time, does providing a reward in the form of a meme or gif to those who completed the survey increase their survey completion and/or active tasks on the next day compared to not providing a reward?

Secondary research questions include

3. Is there an overall effect of Reminder Notification? On average across time, does delivering a push reminder notification at the beginning of the data collection period with a persuasive message result in increased completion of the daily survey and/or active tasks on that same day compared to a push reminder notification without a persuasive message?

4. Is there an overall effect of Post-Active-Task Reinforcement? On average across time, does providing a life insight to those who completed the active task increase survey and/or active task completion on the next day compared to not providing a life insight?

Additional exploratory analyses were planned with the objective of understanding whether effects varied over time and whether observations of context, such as weekend/weekday or rating of a meme or life insight, moderated effects.

**Randomization**. For all four components, the randomization probabilities were 0.50 for deploy notification and 0.50 for do not deploy notification.

**Case Study 3: BariFit**

Individuals who have undergone bariatric surgery should closely monitor their weight and dietary intake while they progress through the stages of recovery. Because physical activity levels predict long-term weight loss and maintenance (Bellicha, Ciangura, Poitou, Portero, & Oppert, 2018; Egberts, Brown, Brennan, & O'Brien, 2012), regular exercise is recommended. However, engaging in these multiple health behaviors can be challenging for an individual attempting to recover from surgery. BariFit is a digital intervention being developed by scientists at the Kaiser Permanente Washington Bariatric Surgery Program to provide low-burden lifestyle change support to facilitate ongoing weight loss. The distal outcome was achievement and maintenance of weight loss after bariatric surgery. As will be shown below, the BariFit optimization trial includes both micro-randomized and baseline-randomized factors and, therefore, is a hybrid of the classical factorial experiment and the MRT.

**Intervention Components and Component Options.** Four components of BariFit were examined in the optimization trial. The first two, Rest Days and Adaptation Algorithm, pertain to adaptive daily step goals. Part of the BariFit intervention involved texting a suggested step goal for the day to each participant each morning to provide guidance for progressively increasing physical activity. The goals were adaptive and based on participants' activity levels over the previous ten days. The Rest Days component had two options: to have a day without a step goal on average one day per week, or to have no rest days and receive the goal every day. The Adaptation Algorithm component concerned how the suggested step goal was computed each day. Based on a prior study, the investigators hypothesized that more variability over time in the step goal suggestions may lead to a higher number of steps per day. The options of this component were two different adaptation algorithms based on a participant's recorded daily step count over the previous ten days: one, the fixed percentile algorithm, provided less variability in the goal suggestions, and the other, the variable percentile algorithm, provided more.

The remaining two components were Activity Suggestions and Reminder to Track Food. The Activity Suggestions component was similar to that described above in HeartSteps, except that the suggestions were delivered via text messages instead of smartphone notifications. The content of the text message suggestion was tailored to the time of day, day of the week, and that day's weather conditions at the participant's home location. Due to the use of text messaging, BariFit did not have access to individuals' location, so messages could not be tailored based on location. As in HeartSteps, there were three component options for the Activity Suggestions:

walking suggestion; anti-sedentary suggestion; or no suggestion. The Reminder to Track Food component was included because tracking food intake is an important predictor of successful weight loss trajectory after surgery (Harkin et al., 2016). The Reminder to Track Food component consisted of a text message, delivered at the start of the day, reminding participants to record their food intake. This component had two options: send the reminder text message or do not send the text message.

**Decision Points**. The Adaptation Algorithm and Rest Days components have one decision point at the beginning of the use of the intervention. For Contextually Tailored Activity Suggestions, there were five daily decision points, pre-specified by participants as times they thought they would be most likely to have opportunities to be physically active. For Reminder to Track Food, there was one decision point every morning.

**Observations of Context and Availability Conditions**. Observations of context were collected primarily for use in subsequent data analysis. Variables included time of day, day of the week, daily weather conditions at the home location, and prior and current step counts. There were no availability conditions. In this study there was no real-time detection of current activity level or whether the individual might be operating a vehicle. Furthermore, the activity suggestions and reminders were delivered via text message, which then remained on the participant's phone indefinitely and could be attended to at the participant's convenience.

**The BariFit Optimization Trial.**

**Factors in the Experiment**. The 120-day BariFit optimization trial involved four factors, one corresponding to each of the four intervention components described above. For all four factors, the factor levels corresponded to the component options described above; thus, the Rest Days, Adaptation Algorithm, and Reminder to Track Food factors had two levels, and the Activity

Suggestions factor had three levels. This trial used a hybrid experimental design that included two baseline-randomized factors, in which randomization occurred once at the outset, and two micro-randomized factors. The Rest Days and Adaptation Algorithm factors were baseline-randomized, and the Contextually Tailored Activity Suggestions and Reminder to Track Food factors were micro-randomized.

The decision about whether to use baseline randomization or micro-randomization for a particular factor depends on the research question being addressed. For the Rest Days and Adaptation Algorithm components, the research question concerned which strategy for delivering a time-varying treatment produced the better outcome; here the strategy was used from the beginning and implemented in the same manner across the entire study. For the Rest Days factor, the investigators wanted to learn whether a strategy that involved having an occasional rest from receiving the daily goal suggestion, as opposed to receiving the suggestion daily, would result in a higher step count across the entire four-month study. Because the two strategies were fixed across the entire study—in other words, a participant either received the suggestions daily across the entire study or had an occasional rest day across the entire study—baseline randomization was called for. For the Adaptation Algorithm factor the research question concerned comparison of two different JITAIs for step goals. Each JITAI is a level for the Adaptation Algorithm factor. Thus participants were randomized at baseline between the two levels.

**Measures of Proximal Outcomes**. For the Adaptation Algorithm and Rest Days components, the proximal outcome was average daily step count across the 120-day study. For Contextually Tailored Activity Suggestions, the proximal outcome was number of steps participants took in the 30 minutes following randomization. For Reminder to Track Food, the proximal outcome was use of the Fitbit application to record food intake at any time on that day.

**Research questions**. The research questions motivating the BariFit MRT were as follows:

1. Is there an overall effect of Adaptation Algorithm? Does delivering a step goal computed using a variable percentile algorithm result in a greater average daily step count, compared to the fixed percentile algorithm?

2. Is there an overall effect of Rest Day? Does including a weekly rest day, on which participants do not receive a step goal, result in a greater average daily step count during the study, compared to not including a rest day?

3. Is there an overall effect of Contextually Tailored Activity Suggestions? On average across time, does delivering a text message with an activity suggestion tailored to the user's context increase physical activity in the 30 minutes after the suggestion is delivered compared to no suggestion?

4. Is there an overall effect of Reminder to Track Food? On average across time, does delivering a reminder result in self-monitoring of food intake that same day compared to no reminder?

In addition, exploratory analyses were planned to examine how contextual variables, such as time of day or day of the week, might moderate any observed effects.

**Randomization**. Randomization for the Adaptation Algorithm and Rest Days factors occurred once, before the start of the experiment, using randomization probabilities of 0.5 for each factor level. For Contextually Tailored Activity Suggestions the randomization probabilities were walking suggestion, 0.15; anti-sedentary suggestion, 0.15; no suggestion, 0.70. For Reminder to Track Food the randomization probabilities were 0.5 for each of the two factor levels. These probabilities, selected to minimize participant burden and avoid habituation, resulted in an average of two text-message pushes per day (adaptive step goals, when present, were sent at the same time as the Track Food Reminder).

## Considerations When Planning and Designing an MRT

In this section we discuss a variety of considerations that go into planning and designing an optimization trial that involves micro-randomization, using the case studies as examples.

**Importance of a Conceptual Framework**

Creation of a scientifically sound and well-specified conceptual model of an intervention is an essential foundation for selection of both the intervention components and their respective proximal outcomes (Collins, 2018; Nahum-Shani et al., 2018). Evaluation of a component in terms of a proximal outcome rests on the assumption that success in affecting the proximal outcomes will translate into success in affecting the distal outcomes. In other words, digital interventions, like most interventions, are based on mediation models, in which proximal outcomes mediate the effect of the intervention components on distal outcomes. The idea is that these proximal outcomes either directly affect the distal outcome (e.g., Contextually Tailored Activity Suggestions lead to increased activity in the form of steps, which leads to weight loss) or form part of a causal chain in which proximal outcomes affect subsequent proximal outcomes until the distal outcome is reached (e.g., Reminder to Track Food leads to tracking food intake, which leads to better control of caloric intake, which leads to weight maintenance). Therefore, the conceptual model must articulate all hypothesized mediated paths.

Note, however, that it is possible for an intervention component to be effective at changing its intended proximal outcome, yet this change in the proximal outcome may not lead to a desired change in a distal outcome. This can happen for a number of reasons, including that the hypothesized causal path was incorrectly specified, that the achieved effect on the proximal outcome is too weak to alter the distal outcome, or that the change in the proximal outcome led to some form of compensatory behavior (e.g., a person who walked in response to Activity Suggestions walked less at other times) that offset its effect on the distal outcome.

**Deciding Which Components to Examine Experimentally**

An investigator designing an MRT can broadly define the term "intervention component" to suit the research questions at hand (Collins, 2018). In both BariFit and HeartSteps, all of the intervention components were designed to have a health benefit, whereas in SARA the intervention components were all strategies to improve engagement in data collection. Note that an intervention component might represent any aspect of an intervention that can be separated out for study, such as the delivery mechanism (e.g., delivering a message via a notification on smartwatch or via a SMS text).

The case studies demonstrate that when conducting an optimization trial, it is not always necessary to examine every component. Some may be considered necessary to implement the rest of the intervention. Examples include components that provide foundational information or maintain interest or engagement in the intervention. Others may already be supported by a sufficient body of empirical evidence or represent current standard of care, so that further experimentation is unnecessary. Such components may be treated as constants in the optimization trial; that is, they are provided to all participants in the same manner. For example, in the SARA MRT, the aquarium environment was a constant component. A constant component may in fact be a JITAI; the aquarium environment is adaptive—badges and rewards are adapted to the participant's adherence over time. When constant components are included in an optimization trial, any results concerning experimental components are conditional on the presence of the constant components. Therefore, it is necessary either to assume that any constant components do not interact with the experimental components, or to consider any constant components as "givens" in the intervention.

**Approach to Randomization**

A decision requiring careful consideration on the part of the investigator is whether a particular intervention component should be examined via micro-randomization or baseline randomization. As the BariFit case study illustrates, the MRT and the factorial experiment are not mutually exclusive; an optimization trial can use hybrid designs that include a mix of micro-randomized factors and baseline randomized factors. Each of these forms of randomization addresses different kinds of research questions.

The motivation for micro-randomizing an intervention component is to gather information needed to optimize the design of a JITAI component. For example, the investigator may wish to assess whether specific options of a component are more effective in some contexts (where context includes recent exposure to the same or other push components), while other options are more effective in other contexts. Micro-randomization is suitable only for a component for which the goal is to develop JITAI component. By contrast, baseline randomization maybe used for all types (JITAI, non-adaptive, time-varying, non-time-varying) of components. Indeed, baseline randomization of JITAI components can make practical sense if the investigator is trying to choose between two well defined JITAI options for a component. For example, recall that the Adaptation Algorithm component of BariFit involved two options. The two options are both JITAIs that differ with respect to how the treatment would be varied across time. Scientific interest lay in ascertaining which of the two pre-specified, fixed decision rules for adapting step goals over time was more effective, not in developing the decision rules. Thus the Adaptation Algorithm component was randomized at baseline. Unlike micro-randomization, baseline randomization is not intended to enable causal inferences about how the relative effects of intervention options vary over time and/or vary by time-varying context.

Once an investigator has decided to use micro-randomization with a particular factor, it is necessary to identify how often randomization can occur and to determine the randomization probabilities. Taken together, these are an important determinant of participant burden. To obtain the most helpful scientific information, the investigator should do everything possible to ensure that the level of burden associated with being a participant in the MRT does not appreciably exceed that associated with the final design of the JITAI. For example, consider an MRT in which the two levels of a factor are deliver a reminder to practice deep breathing versus do not deliver, and the decision points are at each hour of the day. In this setting, using a randomization probability of 0.5 would mean that over each 12-hour day a participant can be expected to receive 6 reminders. If this presents a level of burden or risk of habituation beyond what would be expected in the final design of the JITAI, then this dose of intervention is inappropriate. Moreover, such an approach could lead to dropout from the MRT, which, in turn, can result in biased analyses and a loss of power.

In contrast to other optimization trial experimental designs such as the factorial, in which randomization probabilities are typically kept equal across all levels of a factor (i.e., if there are two levels, probabilities of 0.5 are used), in an MRT randomization probabilities often differ across levels of a factor. This is because thoughtful selection of the randomization probabilities assigned to each level of a factor is one way to minimize burden and habituation. For example, as discussed above, in BariFit the expectation for the Activity Suggestions component was that participants would tolerate approximately 1.5 activity suggestions per day. To achieve this rate, randomization probabilities of .15 were used for each of the two activity suggestions and .7 for the option of no suggestion. On the other hand, the two options for the Reminder to Track Food component were randomized with probability 0.5 because an average of one reminder over each two-day period was seen as tolerable.

**How MRT Design Can Impact Intervention Design**

Any specified availability conditions built into the design of the MRT will constrain the design of the resulting JITAI. Recall that availability conditions are contexts in which participants should not be sent a push notification because of concerns that a push would be potentially aggravating, mistimed, dangerous, or burdensome. The activity suggestions component in the HeartSteps intervention highlights the importance of identifying appropriate availability conditions. If the design of an MRT specifies that an individual is considered unavailable for a particular intervention component in a particular context, then the experimental data will not provide information on the effect of any other option in this context. It follows that the decision rules in the JITAI developed based on this MRT will provide only the do-nothing option in this context. For example, in an MRT conducted for the purposes of increasing engagement in a commercial wellness app (Bidargaddi et al., 2018), burden was at the forefront of the investigative team's considerations, and any participants who had not responded to a certain number of recent prior engagement pushes were considered unavailable for any further engagement pushes. As a result, any decision rules informed by this MRT automatically incorporate this burden consideration.

Which and how many decision points are selected for randomization in an MRT also may have an impact on the design of the intervention. Sometimes it is not necessary to use an MRT to establish the time of a decision point. For example, in SARA the decision point for the Reciprocity Notification component was daily at 4 pm, as adolescents would likely be out of school by then and this time is prior to the data collection period. Existing data can be informative in identifying decision points; in HeartSteps and BariFit, this approach was used to identify time points at which adults might be more responsive to an activity suggestion. Sometimes, however, there are neither natural decision points nor indications from existing data. In this case it may make sense to establish decision points as frequently as possible for the purpose of the MRT, paired with low randomization probabilities to keep the overall number of provided interventions manageable (recall that frequent decision points do not necessarily mean frequent intervention delivery). Then, the resulting data can be analyzed to inform selection of a subset of decision points for the intervention. When designing an MRT it is important to maintain balance between the sometimes competing objectives of obtaining as much scientific information as possible by selection of the frequency of decision points, the randomization probabilities and keeping participant burden at a realistic and sustainable level.

**Measurement of Outcomes**

As the case studies illustrate, in an MRT the components are typically evaluated in terms of time-varying proximal outcome variables. Different components in an intervention will likely target different proximal outcomes, even though the distal outcome is the same for all components in a particular digital intervention. Sometimes the proximal outcome is a short-term measure of the distal outcome. For example, in SARA the distal outcome was overall survey and active task completion during the 30-day study. The proximal outcomes were short-term measures of survey and/or active task completion. For two of the components, this was completion on the same day, and for the remaining two this was completion on the next day. Other times the proximal outcome is not a short-term measure of the distal outcome, but a different variable entirely. In BariFit the distal outcome is weight loss, but the proximal outcome for the Activity Suggestions component is the number of steps participants took in the 30 minutes following randomization, and the proximal outcome for the Reminder to Track Food component is use of the Fitbit application to record food intake. Because in an MRT the effectiveness of intervention components is typically expressed in terms of measures of impact on proximal outcomes, different components can be evaluated in terms of different outcomes, which represent the mediators through which those components are hypothesized to influence the distal outcome.

In any MRT it is necessary to determine not only how each outcome will be measured, but when. If several components are being examined in a single MRT, this may differ across components. It is necessary to select the timing of measurement of each outcome carefully because effect size can vary over time. For example, in HeartSteps the Activity Suggestions component was expected to have its greatest effect in the 30 minutes immediately following the prompt, whereas the Planning Support component was expected to have its greatest effect over the next 24 hours. Choosing the time frame for measuring the proximal outcome in an MRT can be challenging and requires careful thought because a poor choice of timing of outcome measurement has consequences for the scientific results. MRTs are conducted as individuals go about their lives, and the complexities and contingencies of life can introduce noise. If an outcome is measured too early, the effect may not yet have reached a magnitude that is detectable against this noisy background. If it is measured too late, any effect may have decayed to an undetectable level. In either case, the investigator may mistakenly conclude that an effective component was ineffective. It should be noted that the general issue of measurement timing is not specific to MRTs; it arises in all longitudinal research, even panel studies (Collins, 2006; Collins & Graham, 2002).

Decisions about the timing of measurement in the case studies reported here were based primarily on domain expertise. Although behavioral theory could help inform such decisions, at this writing it is largely silent on behavioral dynamics, such as the timing and duration of effects on time-varying variables. More detailed, comprehensive, and sophisticated theories about behavioral dynamics, informed by empirical intensive longitudinal data, are urgently needed in behavioral science. Until such theories are available to provide guidance, we recommend measuring the proximal outcome as close to the delivery of a component, as often, and for as long a duration as is reasonable without being overly burdensome; for example, in HeartSteps a minute level step count is obtained, enabling exploratory analyses examining the choice of 30-minute duration for the proximal outcome. Frequent measurement affords the best chance of observing time-varying effects when they are at their peak.

**Considerations When Sizing an MRT**

When planning any experiment, it is necessary to identify which research questions are primary and which are secondary, and then make the primary research questions the priority when sizing the study. The case studies illustrate that sometimes a research question directly addressed by one of the factors in the experiment is considered secondary. For example, in Heart Steps, the question of whether the factor Planning Support has an overall effect is considered secondary. In many traditional factorial designs, power is identical for all factors with a given expected effect size, making it common for all factors to correspond to primary research questions. By contrast, it is not unusual for power to vary considerably among the factors in an MRT. Thus, when planning an MRT with multiple factors, it is often convenient to size the study based on one or two primary research questions and consider the remaining research questions secondary. For detailed information on power, sample size calculation and MRTs, see (Liao et al., 2016).

## Discussion

**MRTs and the Meaning of Optimization**

MRTs fit naturally within the multiphase optimization strategy (MOST; e.g., Collins, 2018), a framework for development, optimization, and evaluation of behavioral, biobehavioral, and biomedical interventions. Collins (2018) has defined intervention optimization as "the process of identifying an intervention that provides the best expected outcome obtainable within key constraints imposed by the need for efficiency, economy, and/or scalability" (p. 12). The precise constraints that are relevant vary widely across implementations of MOST. In optimization of JITAI components, the key constraints typically are centered on efficiency, which Collins defined as "the degree to which the intervention produces a good outcome while avoiding wasting money, time, or any other valuable resource" (p. 14). Here efficiency primarily means conserving participant time, energy, and attention and minimizing intrusiveness and burden. An efficient JITAI component has a detectable effect in the desired direction, while demanding the least of participants.

MOST is made up of three phases: preparation, optimization, and evaluation. The optimization phase includes one or more optimization trials that are conducted to assess the performance of components and component options. This information is used to choose the best components and component options and to eliminate those that perform poorly. The term optimization trial does not refer to a single experimental design; instead, any of a wide variety of experimental designs may be used for an optimization trial, including, in addition to the MRT, the factorial (e.g., Collins, 2018); fractional factorial (e.g., Collins, 2018); sequential, multiple assignment, randomized trial (SMART; Nahum-Shani et al., 2012); and system identification experiment (Rivera, Hekler, Savage, & Downs, 2018). The selection of the design of the optimization trial, like the selection of the design of any experiment, is driven by the nature of the scientific questions to be addressed and the level and type of resources available to support experimentation. Once the optimization phase of MOST has been completed, the investigator may move to the evaluation phase, in which the performance of the digital intervention involving JITAI components is compared to that of a suitable control treatment in an RCT.

**The Efficiency of MRTs**

MRTs offer considerable efficiency for two reasons. First, because each individual is repeatedly randomized, statistical tests can trade bias and variance to test for treatment effects based on a combination of between-person contrasts and within-person contrasts. This usually enables statistical power to be maintained using far fewer subjects than would be needed in a completely between-subjects experiment. Second, as the case studies illustrate, MRTs can be (although are not necessarily) used to manipulate multiple factors simultaneously, enabling examination of several components in one efficient experiment. In this case, just as in the traditional factorial experiment, a given level of statistical power can be maintained with a much smaller sample size than would be required if a separate individual trial were conducted to examine each component (Collins, Dziak, & Li, 2009).

**Limitations and Future Directions**

More work is needed to integrate MRTs into the general MOST framework. In particular, in this article we discussed the role of the MRT in optimization of decision rules for individual components of a digital intervention. However, in MOST the ultimate goal is optimization of the intervention as a whole, rather than optimization of individual intervention components (although the latter may be a useful step along the way). This is because the costs and effects associated with individual intervention components may not be strictly additive. For example, there may be economies of scale that provide cost savings when two components are delivered together; or the combined effect of two components may be less than the sum of their individual effects. Further research is needed to determine how best to use the results of an MRT in optimization of the whole intervention.

One principle of the MOST framework is continual optimization (Collins, 2018), which states that optimization is an ongoing process of continual improvement of interventions. In the ever-changing digital environment, particularly when the intervention goes to scale, continual optimization on a rapid timetable is essential. One way to accomplish this would be to conduct MRTs on one or more experimental components in a digital intervention in deployment, in much the same way experimental items are included in each graduate record exam (Educational Testing Service, 2017) to inform development of future exams. As new knowledge is gained, the digital intervention will incrementally improve, and updated versions can be pushed out to users. We see this as an intriguing idea that has the potential to maintain and increase the effectiveness of a digital intervention in an efficient and economical manner.

In the introduction to this article we mentioned that MRTs are predominantly used to examine push intervention components, but they could be used to examine pull components as well. For example, consider the setting in which an individual requests content to help manage a cigarette craving; in this case the device could respond in a variety of ways, such as providing different ordered lists of strategies. It might be useful to experiment with the different orderings of the list so that individuals can more quickly access a strategy that is useful in their current

MRT FOR DEVELOPING DIGITAL INTERVENTIONS: DESIGN        38context. In this case, the decision point for the intervention is the user's request for craving strategies, which then, in an MRT, leads to the randomization of the order in which those strategies are presented. Thus investigators may wish to consider using MRTs to help identify how pull content should be provided to which participants under what circumstances.

In the three case studies presented here, the objective of the study was to inform the development of decision rules; once formed these decision rules would be constant across individuals. Thus, although the intervention options delivered to different individuals at different times and in different contexts varies, the way the decision is made about which intervention option to deliver is identical for all participants. An exciting future direction is personalized interventions, in which the decision rules are person-specific. Personalized interventions have the potential to be highly engaging, responsive, and effective. Currently, methods for developing personalized interventions are being developed in the reinforcement learning field (Liao, Greenewald, Klasnja, & Murphy, 2019; Zhou et al., 2018).

**Conclusions**

Digital interventions, which offer the potential to reach unprecedented numbers of individuals with convenient and engaging behavioral interventions, represent an exciting new direction in intervention science. The MRT is an optimization trial design that is particularly useful for JITAIs because it operates in, and takes advantage of, the rapidly changing environments in which JITAIs are implemented. The MRT fits well within the MOST framework, which calls for conducting one or more optimization trials to obtain the information needed to optimize an intervention prior to evaluation. In this article, we reviewed three case studies to illustrate a number of considerations that arise when planning and implementing MRTs. MRTs are a rigorous and efficient way to gain the scientific information needed to select the right tailoring variables, decision rules, and decision points to make up a JITAI. We hope this article will be a helpful resource for investigators who are developing digital interventions that involve JITAIs.

*Figure 1.* Schematic of randomization for the Activity Suggestions factor in the Heart Steps micro-randomized trial (MRT). In each of the 42 days of the experiment, at each prespecified time of randomization, $t_m$, where $m=1$ to 5, an assessment was made of whether the intervention was disabled, or the participant was driving or walking. If any of these was "yes," no randomization was performed. Otherwise, the individual was randomized to be shown a walking activity suggestion ($p=.30$), anti-sedentary suggestion ($p=.30$), or no suggestion ($p=.40$).

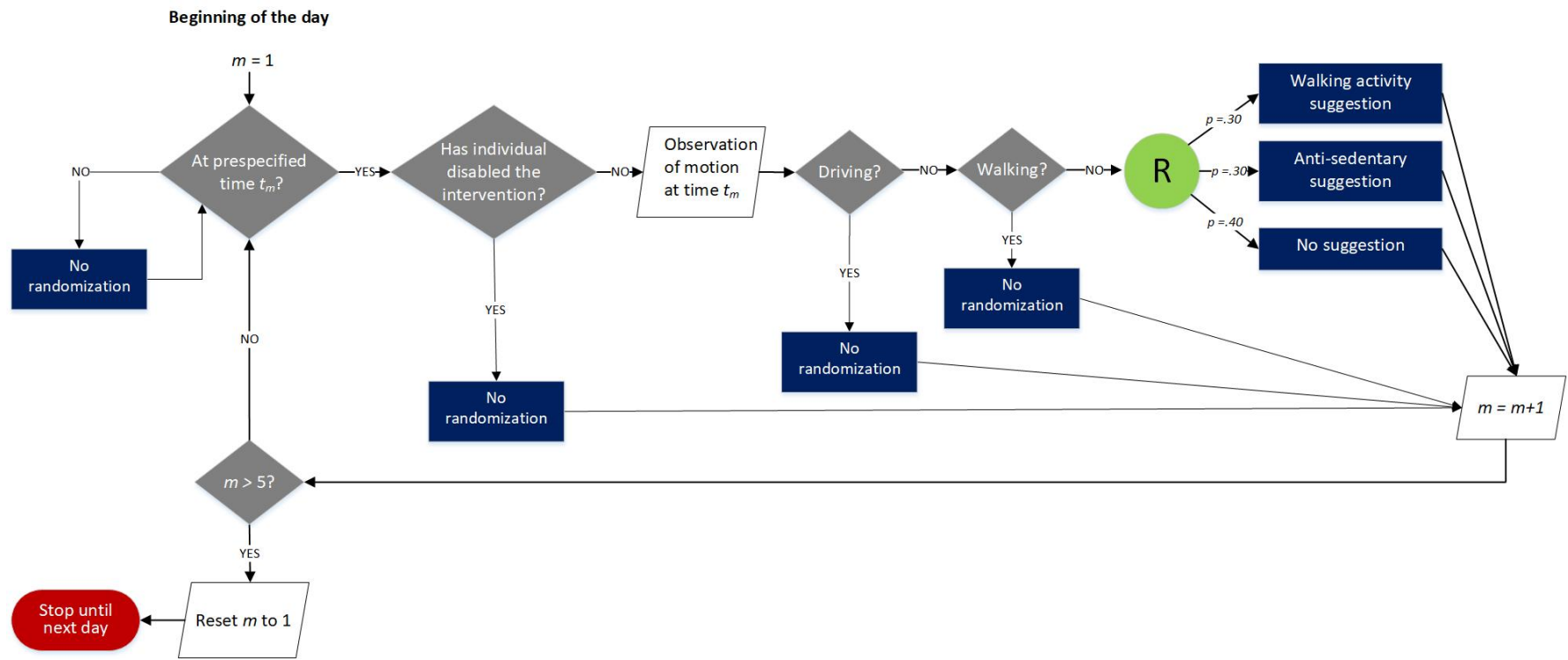